\documentstyle[epsf]{l-aa}

\begin{document}

   \thesaurus{08.16.7} 
   
\title{Time-Resolved Optical Observations of PSR 1509-58
\thanks{Based on observations taken at 
the European Southern Observatory, La Silla, Chile}}


\author{A. Shearer \inst{1}  C. M. M. O' Sullivan \inst{2}, A. Golden\inst{2}, P. V. R. Garcia\inst{3}, M. Redfern \inst{2} 
          A. Danks \inst{4} \and M. Cullum \inst{5}}

   \offprints{A Shearer, andy.shearer@ucg.ie}

   \institute{Information Technology Centre, National University of Ireland Galway, Ireland
	\and Department of Physics, National University of Ireland Galway, Ireland
	\and Department of Astronomy, University of Lyons, Lyons, France
	\and STX/Goddard Space Flight Centre, Greenbank, Maryland, US 
	\and European Southern Observatory, Garching-bei-M\"unchen, Germany 
             }

   \date{Received ...; Accepted ...}

   \maketitle
   \markboth{Shearer, O' Sullivan et al: Time-Resolved Optical Observations of PSR 1509-58}{}


\begin{abstract}

Using time resolved 2-dimensional aperture photometry we have
established that the optical candidate for PSR 1509-58 does not pulse.
Our pulsed upper limits ($m_V$ = 24.3 and $m_B$ = 25.7) put severe
constraints on this being the optical counterpart. Furthermore the
colours of the candidate star are consistent with a main sequence star
at a distance of 2-4 kpc. The probability of a chance coincidence with a
normal star and the difficulty of explaining the lack of pulsed emission
leads us to conclude that this object is an intermediate field star.

\keywords{  -- {\bf pulsars: individual:} PSR 1509-58
                
              }
\end{abstract}

\section{Introduction}

Interest in the optical emission from isolated neutron stars (INSs) has
been growing, as recent improvements in detector sensitivity have
enabled these faint sources to be observed. Optical observations of
neutron stars are important for providing an understanding of pulsar
emission mechanisms and allowing direct observations of the energy
spectrum of the electron pair plasma.   To date 6 INSs have been
detected in the optical. Evidence suggests that it is the age and/or the
period derivative of the INS, rather than its period, that determines
the  optical, and indeed multiwavelength, emission from an INS
(\cite{car94a},  \cite{gol95}).

PSR 1509-58 was initially identified by its X-ray emission
(\cite{sew82}) and shortly afterwards a radio signal, with a period of
150ms, was detected (\cite{man82}). Later studies showed that it had a 
large $\dot P$ (1.5 x 10$^{-12}$ ss$^{-1}$ ), in fact the largest known.
The pulsar has had its second period derivative measured giving a
breaking index ($n=\omega {\ddot \omega}/{\dot \omega^2}$  ) of 2.83
$\pm 0.03$ (\cite{man85}) in close agreement with the expectations of a
radiating dipole (n=3). Its magnetic field ($\propto (P \dot P)^{1/2}$)
is the largest known and its age 1,600, second only to the Crab pulsar
in youth. Its age and location make its association with SNR MSH 15-52
(\cite{sew84}) at a distance of 4.2 kpc (\cite{cas75}) likely, although
this position has been challenged by \cite{str94}, who has proposed a
greater distance, 5.9 kpc, based upon radio dispersion and x-ray spectra
of the extended emission around the pulsar.  It has also been observed
in soft gamma-rays (\cite{gun94}) and a tentative optical counterpart
has been proposed with $m_V \approx 22$ (\cite{car94b}). When taken at a
distance of 4.2 kpc the optical observations indicate an absolute
magnitude of M$_V$$\approx 4.9$ (including the effects of interstellar
extinction), fainter than the Crab pulsar. However, it is much brighter
than would be expected from phenomenological models which have been
successful in describing the X-ray emission from PSR 1509-58 (Pacini and
Savati 1987). This over-luminosity makes its behaviour similar to the
older optical pulsars, PSR0656+14 (\cite{shear97a}) and Geminga
(\cite{shear97b}). Observations of any pulsed optical component will be
crucial in determining what fraction of the emission is thermal and what
is magnetospheric. Only the magnetospheric emission would be expected to
scale in a manner analogous to that described by Pacini and Savati.
Indeed, by considering plausible emission mechanisms (Lu et al 1994) and
high energy observations (\cite{gol95}), it would be reasonable to
expect that most of the emission would be non-thermal.

Given the  importance of a correct determination of the optical emision
from PSR1509-58, we have made time-resolved observations in an attempt
to confirm its identification and determine the optical pulsed fraction.

\section{Observations}

The observations were made on 1995 February 24--25th using University
College Galway's TRIFFID camera mounted at the Nasmyth focus of the ESO
3.5m New Technology Telescope (NTT) at La Silla, Chile. Both nights were
photometric.  The TRIFFID system consists of a multianode microchannel
array (MAMA) 2-dimensional photon counting detector with a B extended
S-20 photocathode (\cite{tim85}) and a fast  data collection system
(\cite{red93}).  The position and time-of-arrival of each photon are
recorded to a precision of 25$\mu$m and 1$\mu$s, respectively.  Absolute
timing is achieved using a GPS receiver and an ovened 10-MHz crystal. On
the NTT the 1024$\times$256-pixel array had an equivalent spatial
resolution of $0''.13$ pixel$^{-1}$. Observations were made with
standard B and V filters. The Crab pulsar and photometric standard stars
were observed each night for calibration  purposes and to ensure that
the system was working correctly. The pulsar observations are summarised
in Table 1. 

\begin{table}
\caption[]{Summary of Observations}
\label{obs}
\begin{flushleft}
\begin{tabular}{lcccc}
\hline\noalign{\smallskip}
 Date   &  UTC  & Filter & Duration &  Seeing\\
 (1995) & \null &  \null & (s)      & \null       \\
\noalign{\smallskip}
\hline\noalign{\smallskip}
Feb 24 & 07:25:43 &  V & 2600   &$ 1".2$ \\
Feb 24 & 08:51:46 &  V & 5100   & $1".2$ \\
Feb 25 & 07:39:49 &  B & 6250   & $1".0$ \\
\noalign{\smallskip}
\hline
\end{tabular}
\end{flushleft}
\end{table}

\section{Data Reduction}

The data were first binned into 1 ms frames and divided by a deep
flatfield taken during the observations. A post-exposure  shift-and-add
sharpening technique was applied to produce an integrated image (\cite{shear96}).  Because of the large telescope aperture ($> 15 r_{0}$,
where $r_0$ is Fried's parameter), this does not produce any significant
improvement in the image above the seeing limit, but corrects for
effects such as telescope wobble.  The optical
candidate was identified in both the integrated B and V images.  Photometry
was carried out using the IRAF daophot package, with the background
level being determined as the mean of the signal in an annulus of 
radius $1''.5$ and width $0''.25$ centred on the candidate position.
 
Photon times were extracted from a window with a diameter equal to the
seeing width centred on the pulsar candidate.  This choice of window
diameter maximises the signal to noise.  The time series was translated
to the solar-system barycentre using the JPL DE200 ephemeris and then
folded in phase using the  PSR 1509-58 ephemeris of (\cite{tay93})
(Table 2). The resulting light curves were   analysed using the
$Z_{n}^{2}$ statistic (\cite{buc89}).

\begin{table}
\caption[]{PSR1509-58 Ephemeris}
\begin{flushleft}
\begin{tabular}{ll}
\hline\noalign{\smallskip}
Parameter & Value\\
\noalign{\smallskip}
\hline\noalign{\smallskip}
$\nu$        & 6.6375697299830 Hz \\
$\dot{\nu}$  & -6.76954$\times10^{-11}$ Hz s$^{-1}$ \\
$\ddot{\nu}$ & 1.96$\times10^{-21}$ Hz s$^{-2}$ \\
Epoch        & 2448355.5 \\
\noalign{\smallskip}
\hline
\end{tabular}
\end{flushleft}
\end{table}

\section{Results}

The optical counterpart ($\alpha_{J2000} = 15^{h}13^{m}55^{s}.52,
\delta_{J2000} = -59^{\circ}8'8''.8 $) proposed by Caraveo et al.
(1994b) can clearly be seen in V (14$\sigma$) and, with less
significance, in B(4$\sigma$).  
 
We find the pulsar candidate has magnitudes $m_{v}=22.4\pm0.4$,
$m_{b}=24.5\pm0.5$. This is to be compared with previous estimates
$m_{v}=22.0\pm0.2$, $m_{b}>23$ (\cite{car94b}).  The
uncertainties in the measurements were  calculated from a combination of
the counting statistics as well as any systematic  error in the
flat-fielding.  Figures 1 and 2 show contour plots of the pulsar
candidate.

\begin{figure}
\epsfysize 3.3truein
\epsffile{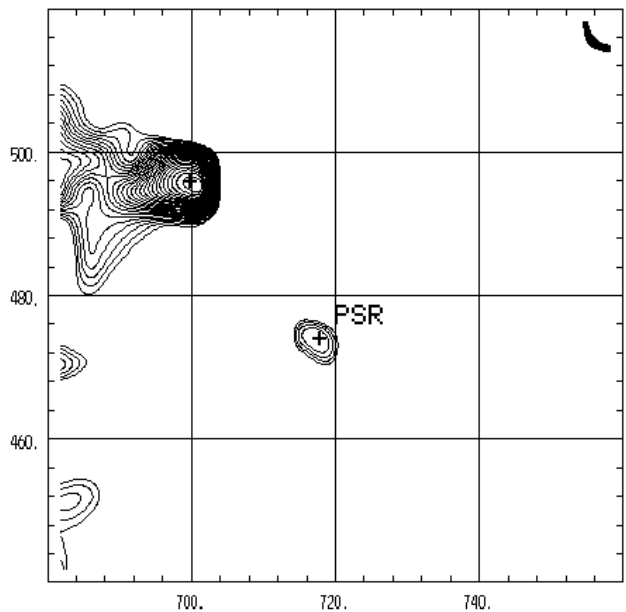}

\caption{128-minute V-band image of the pulsar candidate, labelled PSR.
(1 pixel = $0''.13$)}
\label{vband}
\end{figure}

\begin{figure}
\epsfysize 3.3truein
\epsffile{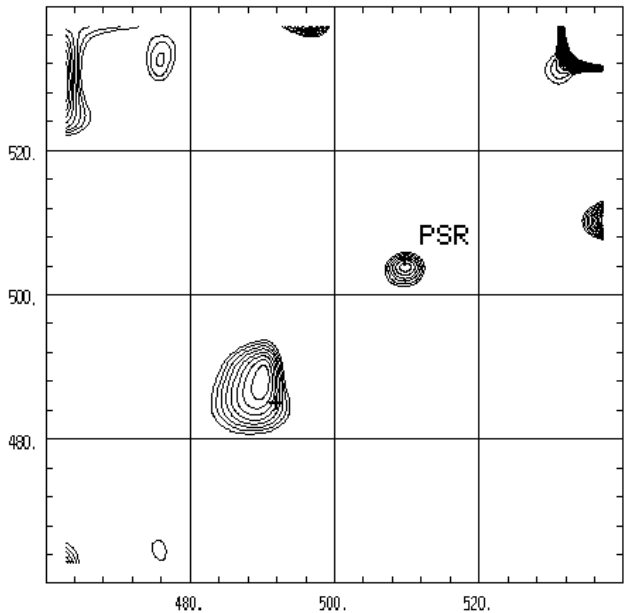}

\caption{104-minute B-band image of the pulsar candidate, labelled PSR.
The crosses correspond to the centres of the stars in the V-band image
which is rotated with respect to this one. (1 pixel = $0''.13$)}
\label{bband}

\end{figure}

No pulsations were seen (to the 1\% significance level) in any of the
time-series  data sets.  An upper limit for the pulsed fraction was
calculated for the B and V emission assuming, conservatively, a duty
cycle in the optical of 50\%. The 3$\sigma$ upper limits were found to
be $m_{v,\rm{pulsed}}>24.3\pm0.2$,  $m_{b,\rm{pulsed}}>25.7\pm0.2$.
These correspond to pulsed fractions $< 18\%$ in V and  $ < 33\%$ in B.
For comparison the Crab pulsar has a pulsed fraction of $>99\%$ and Vela
$ > 77\%$. At the reported pulsar distance of 4.2 kpc ($A_V=4,
E_{B-V}=1.3$), the measured B-V of 2.1 is  equivalent to an intrinsic
colour (B-V)$_{\rm o}=0.8$ and absolute magnitude $M_{v}=5.3$, somewhat
redder and fainter than the Crab pulsar.  At 4.2 kpc this would be
equivalent to a G type main sequence star.

\section{Conclusions}

Our dervived magnitudes and colours, when combined with the lack of
optical pulsations, cast doubt on the Caraveo et al (1994b) candidate
being the optical counterpart of PSR 1509-58. Our data is consistent
with an M type main sequence star at a distance of about 2 kpc.
Alternatively, if it is the pulsar then the pulsed fraction must be
anomolously low (lower than any other optical pulsar) and most of the
radiation thermal. This is contrast with higher energy observations
(dominated by non-thermal emission), where the pulsed fraction increases
with decreasing energy (\cite{gre95}). However if the optical emission
represents the Rayleigh-Jeans tail of the neutron star's black body
spectrum, then with tabulated values for extinction towards MSH15-52,
the distance would have to be $\sim$ 1 kpc assuming a surface
temperature of $3~10^6$ K  in contrast to the expected distance of at
least 4.2 kpc. The presence of a neutron star atmosphere (\cite{pav96})
would not change this distance sufficiently to explain the optical
excess. If the radiation is non-thermal then it is difficult to imagine
a geometry and a mechanism which would give such an anomolously low
pulsed fraction and still be consistent with high energy observations.
Even at the distance derived from radio dispersion (5.9 kpc) the star is
still too red to be explained by a thermal extrapolation alone. A final
answer to the nature of this star will come from spectroscopy - its
magnitude is well within the capabilities of the VLT.


\begin{acknowledgements}

Goddard Space Flight Centre and the European Southern Observatory are
thanked for the provision of their MAMA detectors.  Peter Sinclair and
the ESO detector workshop at La Silla are thanked for their invaluable
help during the observations. The support of FORBAIRT, the Irish
Research and Development agency, is gratefully acknowledged.  M\'{\i}che\'{a}l
Colhoun and Peter O' Kane of University College Galway are thanked for
their technical assistance. CMMO'S is supported by Forbairt under their
Presidential Post-doctoral scheme.

\end{acknowledgements}

\end{document}